\documentclass[journal]{IEEEtran}

\usepackage{cite}
\usepackage{url}
\usepackage{graphicx}
\usepackage{color}

\usepackage{amsmath}
\usepackage{amsthm,amsfonts}
\usepackage{pgfplots}

\usepgfplotslibrary{colorbrewer}

\usepackage[pdfusetitle]{hyperref}
\hypersetup{
    colorlinks,
    linkcolor={red!50!black}, 
    citecolor={blue!50!black},
    urlcolor={blue!50!black}
}

\providecommand*{\mat}[1]{\mathbf#1}
\providecommand*{\M}[1]{\mathbf#1}

\providecommand*{\mrm}[1]{\mathrm{#1}}

\renewcommand{\vec}[1]{{\boldsymbol#1}}
\providecommand*{\V}[1]{\boldsymbol#1}
\providecommand*{\UV}[1]{\hat{\boldsymbol#1}}
\providecommand*{\T}[1]{\mathrm{#1}}
\DeclareMathAccent{\ring}{\mathalpha}{operators}{"17}

\providecommand*{\eu}{\ensuremath{\mrm{e}}}

\providecommand*{\ju}{\ensuremath{\mrm{j}}}

\providecommand*{\diff}{\operatorname{d}\!}

\providecommand*{\diffS}{\operatorname{dS}}
\providecommand*{\diffV}{\operatorname{dV}}
\newcommand{\partder}[2]{\frac{\partial#1}{\partial#2}}
\newcommand{\secondpartder}[2]{\frac{\partial^2#1}{\partial#2^2}}
\newcommand{\Tr}{\mathop{\mrm{Tr}}\nolimits}

\newcommand{\norm}[1]{\lVert#1\rVert} 
\providecommand*{\asinh}{\operatorname{asinh}}
\providecommand*{\atan}{\operatorname{atan}}

\newcommand{\R}{\mathbb{R}{}}
\newcommand{\C}{\mathbb{C}{}}

\newcommand{\ie}{\textit{i.e.}\/, }
\newcommand{\eg}{\textit{e.g.}\/, }
\newcommand{\cf}{\textit{cf.}\/, }

\newcommand{\herm}{\text{H}}
\newcommand{\Id}{\mat{1}}

\newcommand{\Na}{N_\mrm{a}} 
\newcommand{\Ne}{N_\mrm{e}} 
\newcommand{\Nc}{N_\mrm{c}} 
\newcommand{\Nr}{N_\mrm{r}} 
\newcommand{\Nh}{N_{1/2}} 
\newcommand{\reg}{\varOmega}
\newcommand{\regT}{\varOmega_\T{T}}
\newcommand{\regR}{\varOmega_\T{R}}

\newcommand{\rv}{\vec{r}}

\newcommand{\TBs}[1]{}

\graphicspath{{fig/}}

\title{
Spatial Degrees of Freedom and Channel Strength for Antenna Systems
}

\author{
Mats~Gustafsson
~and 
Yaniv~Brick
\thanks{This work was supported in part by the Swedish Research Council SEE-6GIA and SSF Sabbatical and the Israel Science Foundation (ISF) grant No. 442/22.}%
\thanks{M. Gustafsson is with Electrical and Information Technology, Lund University, Lund, Sweden, (e-mails: mats.gustafsson@eit.lth.se).
Y. Brick is with the School of Electrical and Computer Engineering, Ben-Gurion University of the Negev, Beer-Sheva, Israel 8410501 (e-mail:ybrick@bgu.ac.il).
}
}

\begin{document}

\maketitle
\begin{abstract}
The number of spatial degrees of freedom (NDoF) and channel strength in antenna systems are examined within a geometric framework. Starting from a correlation-operator representation of the channel between transmitter and receiver regions, we analyze the associated eigenspectrum and relate the NDoF to its spectral transition (corner).
We compare the spectrum-based effective NDoF and effective rank metrics, clarifying their behavior for both idealized and realistic eigenvalue distributions. In parallel, we develop geometry-based asymptotic estimates in terms of mutual shadow (view) measures and coupling strength. Specifically, we show that while the projected length or area predicts the number of usable modes in two- and three-dimensional settings, the coupling strength determines the average eigenvalue level.
Canonical configurations of parallel lines and regions are used to derive closed-form asymptotic expressions for the effective NDoF, revealing significant deviations from the spectral corner in closely spaced configurations. The results illustrate that these are physically grounded. 
The proposed theory and techniques are computationally efficient and form a toolbox for estimating the modal richness in near-field channels, with implications for array design, inverse problems, and high-capacity communication systems.

\end{abstract}

\section{Introduction}
\IEEEPARstart{E}{lectromagnetic} channels operating in the near field are attracting increasing interest due to emerging applications such as high-capacity wireless links, reconfigurable intelligent surfaces, and compact multi-antenna systems~\cite{Cui+etal2022}. In these regimes, classical far-field intuition based on angular resolution is insufficient: wavefront curvature, strong distance dependence, and geometry-specific coupling significantly influence how many independent communication channels can be supported~\cite{Ji+etal2023,Ruiz-Sicilia2023,Bjornson+etal2024,Maisto+etal2021,Hu+etal2018,Puggelli+etal2025}. A central challenge is, therefore, to quantify the number of degrees of freedom (NDoF) and the associated channel strength in a way that is both physically meaningful and computationally tractable~\cite{Bucci+Franceschetti1989,Poon+etal2005,Migliore2006a,Migliore2008,Franceschetti2015,Franceschetti2017,Bucci+Migliore2025,Dardari2020,Gustafsson2025a,Gustafsson2025c,Jensen+Wallace2008,Nordebo+etal2006,Creagh+etal2025,Ozzola+etal2022}.

A common approach characterizes the NDoF through the singular-value or eigenspectrum of a channel operator. In MIMO array models, dominant singular values correspond to usable sub-channels, while rapidly decaying values represent weak, noise-sensitive modes~\cite{Molisch2011}. For continuous electromagnetic systems, this extends naturally to operator formulations based on Green’s functions, where the eigenspectrum of a correlation operator captures modal coupling between regions. In idealized settings, the NDoF is often associated with a spectral corner separating propagating modes from rapidly decaying reactive (evanescent) contributions~\cite{Bucci+Franceschetti1989,Poon+etal2005,Migliore2006a,Migliore2008,Franceschetti2015,Franceschetti2017,Bucci+Migliore2025,Gustafsson2025a,Gustafsson2025c}.

Complementary insight is provided by geometry-based approaches, where the NDoF is predicted from quantities such as surface area and propagation directions. This connects to spectral results such as Weyl’s law~\cite{Weyl1911,Arendt+etal2009} and sampling-theoretic interpretations of electromagnetic fields~\cite{Franceschetti2017}. Recent work indicates that mutual shadow (or view) measures between regions can predict the spectral corner and provide a physical interpretation of near-field spectra~\cite{Gustafsson2025c,Brick+etal2026}. While such estimates are intuitive and computationally efficient, their relationship to spectral metrics such as the effective NDoF~\cite{Yuan+etal2022,Shiu+etal2000} and effective rank~\cite{Roy+Vetterli2007} remains unclear.

In this paper, we formulate the electromagnetic channel using correlation operators and compare the spectrum-based effective NDoF and effective rank metrics with the geometry-based estimates derived from mutual shadow measures. Their behavior is analyzed for both idealized and realistic eigenspectra. For canonical configurations, stationary-phase analysis yields closed-form asymptotic expressions for the effective NDoF, revealing strong deviations from the spectral corner in closely spaced regimes. These deviations are traced to changes in the propagating eigenspectrum, which transitions from nearly flat for well-separated regions to highly uneven with a few dominant modes at small separations, causing the effective NDoF to vanish as distance decreases. The effective rank is found to be somewhat more robust to variations in the eigenspectrum.  The increased spectral variation in configurations with broad propagation-angle spread is explained by combining observations on the eigenspectrum localization and coupling strength.

The main contributions of this work are:
\begin{itemize}
\item Establish a relationship between the average strength of propagating modes and the electromagnetic coupling between regions.
\item Interpret variations in channel strength in terms of propagation directions and geometric visibility.
\item Provide a comparative analysis of effective NDoF and effective rank, clarifying their relation to the spectral corner.
\item Show that the effective rank exceeds the effective NDoF and more closely tracks the spectral corner for electrically large structures.
\item Derive asymptotic expressions for the effective NDoF in canonical geometries and relate them to mutual shadow–based estimates.
\end{itemize}

The remainder of the paper is organized as follows. Section~\ref{S:DoF} introduces the channel and correlation-operator models, defines spectrum-based NDoF metrics, and presents geometry-based estimates. Section~\ref{S:AsymtoticNe} analyzes canonical configurations and derives asymptotic formulas for parallel lines and planar regions. 
Section~\ref{S:ChannelStrength} discusses the propagating eigenspectra. 
Conclusions are presented in Sec.~\ref{S:conclusion}.
Technical derivations and details of the Green's function are collected in the appendices.

\section{Degrees of freedom and channel strength}\label{S:DoF}
In the analysis of communication between two array antennas (e.g., with $N_\T{T}$ transmitting and $N_\T{R}$ receiving antennas), the singular values of the channel matrix $\M{H} \in \C^{N_\T{R} \times N_\T{T}}$ characterize the strength of the available sub-channels~\cite{Molisch2011}. Weak channels require high signal-to-noise ratios to be useful, and the number of usable sub-channels defines the NDoF~\cite{Franceschetti2017}. The squared singular values of $\M{H}$ are the eigenvalues of the $N_\T{T} \times N_\T{T}$ correlation matrix $\M{H}^{\herm}\M{H}$.

For arbitrary array antennas situated in two regions, we construct a communication channel between sources in $\regT$ and fields in $\regR$ based on the appropriate (dyadic) Green's function $\M{G}(\V{r},\V{r}')$. Based on this channel, we form an eigenvalue problem for the continuous correlation operator 
\begin{equation}
    \M{R}(\V{r}_2,\V{r}_1) = \int_{\regR}\M{G}^{\ast}(\V{r}_2,\V{r}')\cdot\M{G}(\V{r}',\V{r}_1)\diffV'
    \label{eq:CorrelationOp}
\end{equation}
over the current density $\V{J}(\V{r})$ in $\regT$
\begin{equation}
   \int_{\regT} \M{R}(\V{r}_2,\V{r}_1)\cdot\V{J}_n(\rv_1)\diffV_1 = \zeta_n\V{J}_n(\rv_2),
   \label{eq:CorrelationOpEig}
\end{equation}
where the superscript $\ast$ denotes the complex conjugate.
This is a compact operator with a finite number of non-negligible eigenvalues $\zeta_n \geq \zeta_0 > 0$ (i.e., above a threshold $\zeta_0$)~\cite{Bucci+Franceschetti1989,Bucci+Migliore2025}. The number of significant eigenvalues scales with the areas of $\regT$ and $\regR$ for surface domains in $\R^3$, and with their lengths for line domains in $\R^2$ ~\cite{Brick+etal2026,Gustafsson2025c}. Note that for surface domains in $\R^3$ and curves in $\R^2$, the eigenvalues $\zeta_n$ 
have a dimension of area. The dyadic Green's function notation in~\eqref{eq:CorrelationOp} is general and, depending on the formulation, may represent the free-space scalar Green's function (for single- or double-layer sources and scalar fields), Green's dyadics for current sources and electric fields, or combinations relating electromagnetic fields to physical or equivalent currents; see App.~\ref{S:Green}.

The results presented in this paper are largely independent of the specific Green’s function. The numerical evaluations are performed using a discretization of the domains with approximately five points per wavelength, and the resulting eigenspectra exhibit negligible dependence on further discretization refinement, see~\cite{Brick+etal2026,Gustafsson2025c} for details. Cases with electromagnetic fields can often be characterized by DoFs of two polarization~\cite{Franceschetti2017,Bucci+Migliore2025}. The NDoF can be reduced for cases with symmetries unless electric and magnetic currents (or single and double-layer sources) are included~\cite{Gustafsson2025a}.    

\begin{figure}
    \centering
    \includegraphics[width=\linewidth]{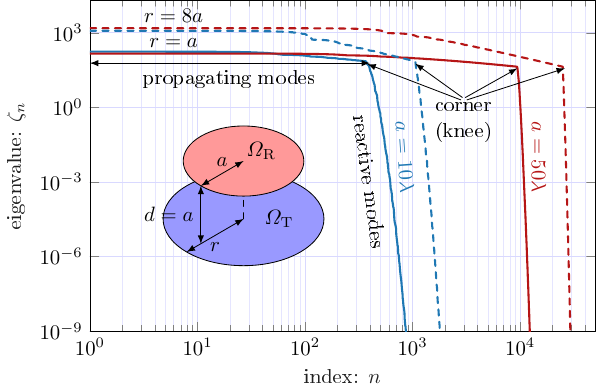}
    \caption{Eigenspectra for the channels between a transmitting disc region $\regT$ with radius $r\in\{1,8\}a$ and receiving disc region $\regR$ with radius $a=\{10,50\}\lambda$ separated by a distance $d=a$. The modes up to and after the corner are termed here ``propagating" and ``reactive", respectively.}
    \label{fig:discspectrumill}
\end{figure}

The eigenspectrum typically exhibits a sequence of nearly constant eigenvalues followed by a region of rapidly decaying ones~\cite{Bucci+Franceschetti1989,Franceschetti2017,Miller2019,Solimene+etal2018,Solimene+etal2019,Bucci+Migliore2025,Maisto+etal2021,Kuang+etal2025}, as illustrated in Fig.~\ref{fig:discspectrumill}. We refer to the modes preceding the transition as propagating and those following it as reactive. The propagating modes are typically associated with smooth current distributions in the region. In contrast, the reactive modes are more strongly associated with increasingly oscillatory currents, such as higher-order modes on spherical region~\cite{Chu1948,Harrington1960}, or currents concentrated at the edges of the region resembling diffraction~\cite {Brick+etal2026}.    

The NDoF is associated with this transition corner (or “knee”), where the spectral behavior changes character. In the idealized case, characterized by a uniform plateau followed by negligible eigenvalues, this corner is well defined and straightforward to identify. In more realistic scenarios, however, the transition may be gradual or involve multiple knees, making a unique definition less clear~\cite{Solimene+etal2013}.
As the electrical size increases, the transition becomes more pronounced, which motivates an asymptotic characterization of the NDoF. In this work, we therefore focus on electrically large configurations with simple geometries in which a dominant corner is present~\cite {Gustafsson2025a,Gustafsson2025c,Brick+etal2026}. This is also similar to the use of asymptotics in Weyl's law~\cite{Arendt+etal2009} and sampling~\cite{Franceschetti2017}.  

As an example, consider the scalar Green's function channel between two discs with different radii, $r$ and $a$ with a spacing $d=a$. The eigenspectrum for this case is shown in Fig.~\ref{fig:discspectrumill}. The eigenvalues for the equal radii case $r=a$ are approximately constant up to the corner, thereafter they decrease rapidly. The corner points are approximately at $n\in\{380,9400\}$ for the sizes $a\in\{10,50\}\lambda$. This is consistent with the classical $\lambda^{-2}$ scaling with the wavelength $\lambda=2\pi/k$ of the NDoF~\cite{Arendt+etal2009,Miller2019,Bucci+Migliore2025}. The $r=8a$ cases are similar, with corner positions at approximately $n\in\{1050,24800\}$. However, here the spread of the propagating eigenvalues is larger, \ie there is approximately a factor of 10 between the first eigenvalues and the eigenvalues just before the corner.

Although these cases are numerically tractable, analytical estimates of the eigenspectrum provide insight. Several techniques have been developed to quantify the NDoF for these and related situations. These include spectrum-based estimators, such as the effective NDoF~\cite{Yuan+etal2022,Shiu+etal2000} and effective rank~\cite{Roy+Vetterli2007}, and geometry-based ones, like those relying on paraxial approximations~\cite{Miller2019,Piestun+Miller2000} or shadow (projected, view) area calculation ~\cite{Gustafsson2025a,Gustafsson2025c,Brick+etal2026}.

\subsection{Spectrum Based Estimates}
Singular value thresholding is commonly used to regularize ill-posed linear inverse problems and numerical approximations~\cite{Hansen2010}. It defines the NDoF as the number of singular values satisfying $\sigma_n \geq \epsilon \sigma_1$, i.e., relative to the largest singular value. For the eigenvalues in~\eqref{eq:CorrelationOpEig}, ordered by decreasing magnitude, this corresponds to $\zeta_n \geq \epsilon^2 \zeta_1$. Alternatively, normalizing by the least-squares norm yields an NDoF defined by $\zeta_n \geq \epsilon^2 \sum \zeta_m$. The resulting NDoF depends on the chosen threshold $\epsilon$. 

Much research has focused on parameter-free techniques to define an NDoF.  
The effective NDoF~\cite{Yuan+etal2022,Shiu+etal2000} is commonly used in wireless communications. It is defined by the quotient
\begin{equation}
    N_\T{e}
    =\frac{(\sum\zeta_n)^2}{\sum\zeta_n^2}
    =\frac{\Tr(\M{H}^{\T{H}}\M{H})^2}{\Tr(\M{H}^{\T{H}}\M{H}\M{H}^{\T{H}}\M{H})}
    =\frac{\norm{\M{H}}^4_\T{F}}{\norm{\M{H}^{\T{H}}\M{H}}^2_\T{F}},
    \label{eq:eNDoF}
\end{equation}
here also expressed using the channel matrix $\M{H}$, with either traces or Frobenius norms. 

The effective rank~\cite{Roy+Vetterli2007} metric, defined by
\begin{equation}
    N_\T{r} = \exp\big(-\sum_n(\zeta_n\ln\zeta_n)\big)
    =\prod_n\zeta_n^{-\zeta_n}
    \label{eq:rNDoF}    
\end{equation}
assuming the normalization $\sum_n\zeta_n=1$, attempts to estimate the rank of a matrix. This metric is less commonly used than the effective NDoF for estimating the NDoF of communication channels.  
For an ideal channel
consisting of $\Na$ identical eigenvalues followed by negligible eigenvalues, the effective NDoF~\eqref{eq:eNDoF} and effective rank~\eqref{eq:rNDoF} coincide with the corner position whereas for more general distributions they typically differ. Jensen's inequality shows that $\Nr\geq \Ne$. 

The eigenvalue sum, which appears in the normalization of~\eqref{eq:rNDoF} and in the numerator of~\eqref{eq:eNDoF}, is sometimes referred to as the coupling strength~\cite{Miller2019,Kuang+etal2025}. For channels modeled by a scalar Green's function, $G=\exp(-\ju k r)/(4\pi r)$ with $r=|\V{r}|$, it is given by
\begin{equation}
 \sum_n\zeta_n=
    \norm{\M{H}}^2_\T{F}
    =\int_{\regT}\int_{\regR} |G(\V{r}-\V{r}')|^2\diffS \diffS'.
    \label{eq:CouplingStrength}
\end{equation}
This term is frequency independent and easily evaluated numerically for arbitrarily shaped regions. For channels modeled via Green's dyadics, it exhibits weak frequency dependence but stabilizes at high frequencies. For the 2D Green's function, the Frobenius norm in the expression is proportional to the wavelength in the high-frequency limit, see App.~\ref{S:Green}.

The corner position $\Nc$ can also be determined by inspection or image processing techniques. For the simple cases in this paper with one dominant corner, the position can 
\eg be estimated by the minimal least squares fit of the spectrum in a logarithmic scale $(\ln n,\ln\zeta_n)$ to two lines.

\subsection{Geometry-Based Estimates}
Geometry-based approaches use simple properties of the regions, such as their surface area or shadow (projected) area, to estimate the NDoF. This resembles the fundamental results on the Laplacian eigenvalue distribution by Weyl~\cite{Weyl1911,Arendt+etal2009} and sampling theorems by Whittaker, Nyquist, and Shannon~\cite{Franceschetti2017}. These results are asymptotic, \eg in $\R^1$ stating that the number of modes~\cite{Weyl1911,Arendt+etal2009} on a line with length $\ell$ is $2\ell/\lambda$, which corresponds to $\lambda/2$ sampling~\cite{Franceschetti2017}.

The total mutual shadow (projected or view) area $A_\T{TR}$ (or length $L_\T{TR}$) between a transmitter and receiver measured in wavelengths determines the asymptotic NDoF~\cite{Gustafsson2025a,Gustafsson2025c}
\begin{equation}
    \Na = \begin{cases}
        L_\T{TR}/\lambda & \text{in } \R^2 \\
        A_\T{TR}/\lambda^2 & \text{in } \R^3 \\
    \end{cases}
    \label{eq:Na}
\end{equation}
per polarization DoF. The mutual shadow area can be evaluated analytically for canonical geometries and numerically for arbitrary-shaped structures~\cite{Gustafsson2025c}. For a simple shape in which every point of $\regR$ is visible from $\regT$, the total mutual shadow (view) area is~\cite{Gustafsson2025c,Brick+etal2026}
\begin{equation}
    A_\T{TR}=
    \int_{\regT}\int_{\regR} \frac{|\UV{n}'\cdot \V{R}|\ |\UV{n}\cdot\V{R}|}{|\V{R}|^4}\diffS' \diffS,
    \label{eq:ShadowArea}
\end{equation}
with $\V{R}=\V{r}-\V{r}'$ denoting a vector connecting points in $\regT$ and $\regR$ and $\UV{n}$ the unit normal of the surface.
In 2D, it reduces to the total mutual shadow length
\begin{equation}
    L_\T{TR}=
    \int_{\regT}\int_{\regR} \frac{|\UV{n}'\cdot \V{R}|\ |\UV{n}\cdot\V{R}|}{|\V{R}|^3}\diff l' \diff l,
    \label{eq:ShadowLength}
\end{equation}
see~\cite{Gustafsson2025c} for more general configurations.
The total mutual shadow (projected) area and length simplify to $\pi A$ and $2L$ for a convex transmitter region surrounded by a receiver region~\cite{Gustafsson2025a}, with surface area $A$ and circumference $L$, in agreement with Weyl's law and sampling theory~\cite{Gustafsson2025a}.

Similar expressions to~\eqref{eq:ShadowArea} and~\eqref{eq:ShadowLength} have been extensively used in thermal radiative heat transfer~\cite{Welty+etal2020,Howell+Menguc2011}, where they are related to radiation view factors. Many analytical evaluations for simple configurations are presented in~\cite{Welty+etal2020,Howell+Menguc2011}. For example, the two discs in Fig.~\ref{fig:discspectrumill} have a total mutual shadow (view) area
\begin{equation}
A_\T{TR}=\frac{\pi^2}{2}\left(
\varDelta-\sqrt{\varDelta^2-4a^2r^2}
\right)
\label{eq:ShadowA2discs}
\end{equation}
with $\varDelta=a^2+r^2+d^2$, giving $\Na\in\{377,9425\}$ for the $r=a$ case and $\Na\in\{972,24289\}$ for the $r=8a$ case using~\eqref{eq:Na}. These are in good agreement with the knee positions seen in the numerical data in Fig.~\ref{fig:discspectrumill}.

For well-separated small regions, the mutual shadow area~\eqref{eq:ShadowArea} simplifies to
\begin{equation}
    A_\T{P}
    =\frac{1}{d^2}
    \int_{\regT}|\UV{n}\cdot \UV{R}|\diffS 
    \int_{\regR}|\UV{n}\cdot \UV{R}|\diffS,
    \label{eq:Aparaxial}
\end{equation}
with the distance $d$ and direction $\UV{R}=\V{R}/R$. For perpendicular surfaces, this expression reduces to the well-known paraxial approximation~\cite{Miller2019} $A_\T{P}=A_\T{T}A_\T{R}/d^2$, where $A_\T{X}$ denotes the area of $\reg_\T{X}$ for $\T{X}\in\{\T{T},\T{R}\}$.

Combining the coupling strength~\eqref{eq:CouplingStrength} with the mutual shadow area~\eqref{eq:ShadowArea} shows that the average level of the propagating part of the spectrum per squared wavelength is approximately 
\begin{equation}
    \frac{(4\pi)^2\sum\zeta_n }{\lambda^2\Na}    =\frac{\displaystyle\int_{\regT}\int_{\regR} \frac{1}{R^{p}}\diffS \diffS'}{\displaystyle\int_{\regT}\int_{\regR} \frac{|\UV{n}'\cdot \UV{R}|\ |\UV{n}\cdot\UV{R}|}{R^p}\diffS \diffS'}
    \geq 1,
    \label{eq:averagepropspec}
\end{equation}
assuming a scalar Green's function~\eqref{eq:Green3D} and $p=2$. This purely geometrical quantity reduces to unity in the paraxial limit for perpendicular surfaces. The corresponding average in 2D is obtained by setting $p=1$ in~\eqref{eq:averagepropspec},
where the high-frequency approximation of the Green's function~\eqref{eq:Green2D} is used. 

The identity~\eqref{eq:averagepropspec} is here referred to as the asymptotic average channel strength. Note that, for arbitrary-shaped regions, the integration in the numerator extends over the entire regions, whereas that in the denominator is effectively restricted to their mutually observable parts~\cite{Gustafsson2025c}, which increases the resulting average channel strength. The factor $\lambda^2$ in~\eqref{eq:averagepropspec} compensates for the dimensionality of $\zeta_n$ (area) in~\eqref{eq:CorrelationOpEig}. For other configurations, such as lines or volumes in 3D, corresponding weightings can be obtained straightforwardly by combining the coupling strength with the appropriate shadow measure.

The asymptotic average channel strength~\eqref{eq:averagepropspec} also suggests a simple threshold-based NDoF estimate $\Nh$ defined by the number of normalized eigenvalues greater than $1/2$, \ie
\begin{equation}
    \Nh=\T{card}\left\{\frac{(4\pi)^2\zeta_n }{\lambda^2} \geq \frac{1}{2}\right\}.
    \label{eq:Nh}
\end{equation}

\subsection{Comparison}

\begin{figure}
    \centering
    \includegraphics[width=\linewidth]{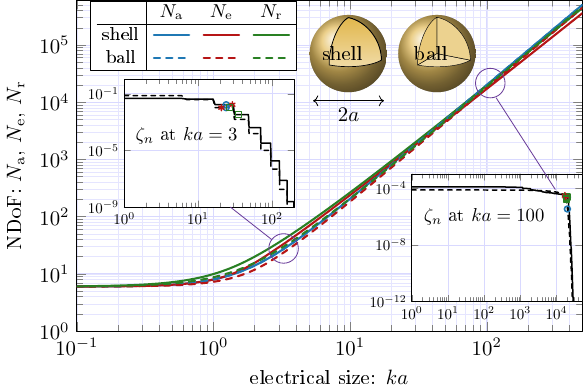}
    \caption{NDoF estimates $\Na,\Ne$, and $\Nr$ for spherical shells and balls versus the electrical size $ka$. Insets depict the corresponding eigenvalues $\zeta_n$ at $ka\in\{3,100\}$ with markers indicating the estimated NDoF.}
    \label{fig:spheigcomp}
\end{figure}
The NDoF metrics $\Ne$ in~\eqref{eq:eNDoF} and $\Nr$ in~\eqref{eq:rNDoF} are first evaluated for spherical shell and solid ball transmitter domains and receivers on the entire far-field sphere. The sources are electric and magnetic currents, for electrical sizes $ka\in[1,500]$, as depicted in Fig.~\ref{fig:spheigcomp}. 
For both geometries, both spectral estimates agree well, converge to 6 for $ka\to 0$, and scale with $k^2$ for large $ka$. Small differences are mainly observed around $ka\approx 1$, but also a slight difference in the slope for large $ka$ can be seen. The comparison also includes a modified NDoF estimate based on the shadow area~\eqref{eq:Na} with a low-frequency asymptotic correction~\cite{Gustafsson2025c}, \ie $\Na\to\Na+6$.

The NDoF in Fig.~\ref{fig:spheigcomp} also agree with the classical approach based on the number of propagating spherical modes~\cite{Chu1948,Harrington1961,Bucci+Isernia1997} 
\begin{equation}
    N = 2L(L+2) = 2L^2+4L
    \label{eq:sphH}
\end{equation}
for spherical harmonic order $L$, using the common estimate $L=ka\geq 1$ for the onset of modes. Here, we also note that $N\approx 2\Na=2\pi A/\lambda^2$ as $\lambda\to 0$ in agreement with~\eqref{eq:Na} and~\eqref{eq:ShadowArea}.
The eigenvalue curves $\zeta_n$ for $ka\in\{3,100\}$ are depicted in the insets in logarithmic scale.  The corners are observed around $2\Na=2(ka)^2=\{18,2000\}$ for the two cases, in agreement with~\eqref{eq:sphH}. The corner becomes increasingly pronounced as $ka$ increases. Until the corner is approached, the spectrum is quite uniform in magnitude.

\begin{figure}
    \centering
    \includegraphics[width=\linewidth]{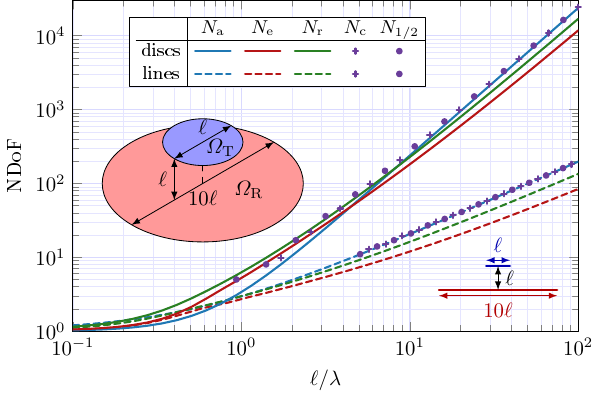}
    \caption{NDoF estimates $\Na,\Ne,\Nr,\Nc,\Nh$ for the channels between two discs (solid curves) in $\R^3$ and two lines (dashed curves) in $\R^2$.}
    \label{fig:DiscsLines}
\end{figure}

As a second example, the various NDoF estimates for channels between two disc domains in $\R^3$ and two line domains in $\R^2$ are shown in Fig.~\ref{fig:DiscsLines} (see inset for geometry details). For small electrical sizes $\ell/\lambda$, the NDoFs estimates are low and differ from one another by $\pm 1$ roughly. In this regime, $\Na$ is modified by including the low-frequency contribution~\cite{Gustafsson2025c}, \ie $\Na \to \Na + 1$, and the corner NDoF $\Nc$ and threshold NDoF $\Nh$ are excluded from the study due to the inherent difficulty of defining a distinct spectral corner when only a few modes are present.

For electrically larger sizes, the NDoF estimates exhibit approximately straight line trends in the log-log plot, indicating a power law behavior
\begin{equation}
    N_\T{x} \approx \alpha_\T{x}/\lambda^p  
    \quad\text{for } N_\T{x}\gg 1
\end{equation}
with $p=2$ for the discs and $p=1$ for the lines. The dominant term $\alpha_\T{x}$ is determined by the short wavelength limit of the NDoF. 
The NDoF estimates in Fig.~\ref{fig:DiscsLines} diverges and are ordered as $\Ne < \Nr < \Na \approx \Nc \approx \Nh$. The inequality $\Ne < \Nr$ holds in general; to further examine the relationship among the remaining estimates, the offsets $\alpha_\T{x}$ between the curves should be analyzed. 

As part of the proposed analysis, in the next section, the small-wavelength asymptotic behavior of $\Ne$ is examined.

\section{Asymptotic evaluation of $\Ne$}\label{S:AsymtoticNe}
The effective NDoF, 
$\Ne$ in~\eqref{eq:eNDoF}, is typically evaluated numerically, as illustrated in Fig.~\ref{fig:spheigcomp} and Fig.~\ref{fig:DiscsLines}. While the results in Fig.~\ref{fig:spheigcomp} show good agreement, those in Fig.~\ref{fig:DiscsLines} exhibit more significant deviations. As shown here, the discrepancies increase for closely spaced configurations. These observations, which rely on numerical experiments, are supported by an asymptotic analysis of $\Ne$ in the electrically large regime, which enables a direct comparison between the $\Ne$ and $\Na$ metrics.

In this asymptotic analysis, $\Ne$ is evaluated by reformulating the terms in~\eqref{eq:eNDoF} as multidimensional integrals and subsequently deriving their leading-order behavior.
The denominator in~\eqref{eq:eNDoF}, \ie $\sum\zeta_n^2$, is 
\begin{equation}
    \norm{\M{H}^{\T{H}}\M{H}}^2_\T{F}
    =\int\!\!\!\int\!\!\!\int\!\!\!\int
    G^{\ast}_{1,1'}
    G_{2,1'}
    G^{\ast}_{2,2'}
    G_{1,2'}
    \diffS_2\diffS_2'\diffS_1\diffS_1'
    \label{eq:GGint}
\end{equation}
with the short-hand notation $G_{m,n'}=G(\V{r}_m,\V{r}_n')$.
This multidimensional oscillatory integral (defined over $\mathbb{R}^4$ for lines, $\mathbb{R}^8$ for surfaces, and $\mathbb{R}^{12}$ for volumes) is not well suited for direct numerical evaluation. However, it is amenable to asymptotic analysis via stationary phase methods~\cite{Bleistein+Handelsman1986}, which yield analytic expressions in the short-wavelength limit.
  
The product of the Green's functions~\eqref{eq:GGint} can be written as $\exp(-\ju k\phi)h$, where the phase $\phi$ is
\begin{equation}
    \phi = |\V{r}_1-\V{r}_1'|-|\V{r}_2-\V{r}_1'|+|\V{r}_2-\V{r}_2'|-|\V{r}_1-\V{r}_2'|
\end{equation}
and $h$ contains a product of $1/|\V{r}_1-\V{r}_1'|$ type terms. Assuming a single stationary point at $\V{r}_0$ gives~\cite{Bleistein+Handelsman1986}
\begin{equation}
    \int f(\V{r})\eu^{\ju k\phi(\V{r})}\diffV
    =\lambda^{n/2} \frac{f(\V{r}_0)\eu^{\ju k\phi(\V{r}_0)+\ju\frac{\pi}{4} \T{sgn}(\M{H}_\phi)}}
    {|\det(\M{H}_\phi)|^{1/2}}
    +o(\lambda^{n/2})
    \label{eq:stationaryphase}
\end{equation}
as $\lambda\to 0$, where $\M{H}_\phi$ denotes the Hessian at $\V{r}_0$.
The stationary points with respect to variations over $\V{r}_2,\V{r}_2'$ are given by the zeros of $\nabla_2 \phi$ and $ \nabla_2' \phi$.

We apply the technique to the canonical cases of two parallel lines and two parallel plates. 

\subsection{Two parallel lines}\label{S:TwoLines}
\begin{figure}
    \centering
    \includegraphics[width=0.95\linewidth]{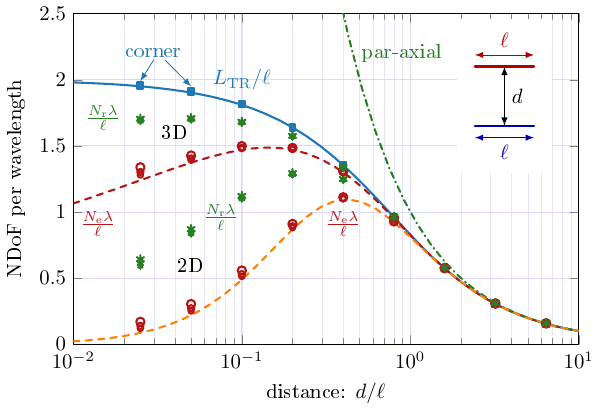}
    \caption{NDoFs per wavelength for the channel between two lines with length $\ell$ and separation distance $d$. The shadow-length estimate~\eqref{eq:ShadowLength2lines} is shown by the solid curve, the asymptotic effective NDoF in 3D and 2D from~\eqref{eq:NeNDoF2lines} and~\eqref{eq:eDoF2lines2D}, respectively, by dashed curves, and the paraxial approximation by a dash-dotted curve. Analytical results are compared with numerical estimates of the effective NDoF~\eqref{eq:eNDoF}, effective rank~\eqref{eq:rNDoF}, and corner position for wavelengths $\lambda\in\{1,2,4\}0.01\ell$, indicated by small to large markers.}
    \label{fig:eDoF2lines}
\end{figure}

The asymptotic effective NDoF for short wavelengths is first determined for two parallel lines in 3D with length $\ell_1$ and $\ell_2$ separated by a distance $d$, as shown in the inset of Fig.~\ref{fig:eDoF2lines}. The stationary phase approximation of the effective NDoF~\eqref{eq:eNDoF} derived in App.~\ref{S:Lines} is
\begin{equation}
    \Ne^{0} = \frac{\ell}{2\lambda}\frac{\big(\ln(1+\beta^2) - 2\beta\atan(\beta)\big)^2}{\beta^2\asinh(\beta)-\beta\sqrt{\beta^2+1}+\beta}
    \label{eq:NeNDoF2lines}
\end{equation}
as $\lambda\to 0$ with $\beta=\ell/d$.  
This asymptotic expression for the effective NDoF is illustrated in Fig.~\ref{fig:eDoF2lines} (dashed red line labeled ``3D"). For large separation distances $\beta\to 0$, the limit  $\Ne^0\to \ell^2/(d\lambda)$ coincides with the paraxial limit.  For short distances, the limit is
\begin{equation}
    \Ne^0\approx \frac{\ell}{2\lambda}
    \frac{2\pi}{\ln(2\beta)}
    \to 0 \text{ as } \beta\to\infty,
\end{equation}
\ie it vanishes logarithmically as $d\to 0$, as suggested by the slow decay of the corresponding curve in Fig.~\ref{fig:eDoF2lines}.

The NDoF based on the total mutual shadow length~\eqref{eq:ShadowLength} (blue line in Fig.~\ref{fig:eDoF2lines}) is given by 
\begin{equation}
    \Na =\frac{L_\T{TR}}{\lambda}
    =\frac{2\ell}{\lambda}\frac{\sqrt{1+\beta^2}-1}{\beta}
    \geq \Ne^0.
    \label{eq:ShadowLength2lines}
\end{equation}
It is greater than or equal to the effective NDoF~\eqref{eq:NeNDoF2lines} for sufficiently short wavelengths and also approaches the paraxial approximation for large separation distances and $2\ell/\lambda$ as $d\to 0$ in agreement with Weyl's law~\cite{Gustafsson2025c}. 

The closed form shadow area estimate and asymptotic effective NDoF are compared with numerical evaluation of the various NDoF estimates, for a model of the channel between the lines obtained by their sampling with five points per wavelength, for $\lambda \in \{1,2,4\}0.01\ell$, as shown in Fig.~\ref{fig:eDoF2lines} (corresponding color markers of sizes decreasing with $\lambda$). The numerically evaluated $\Ne$ shows increasingly good agreement with the analytical result as $\lambda$ decreases. Likewise, the numerically estimated eigenvalue corner points agree well with the mutual shadow length~\eqref{eq:ShadowLength2lines}. The limits diverge as $d/\ell \to 0$, where it is seen that $\Ne\lambda/\ell$ decreases for small $d/\ell$ and falls below the mutual shadow length $L_\T{TR}$. This decrease in effective NDoF at small separation distances is somewhat counterintuitive.   This can be explained by the behavior of the eigenvalue spectrum as a function of $d$, as is shown next. 

\begin{figure}
    \centering
    \includegraphics[width=\linewidth]{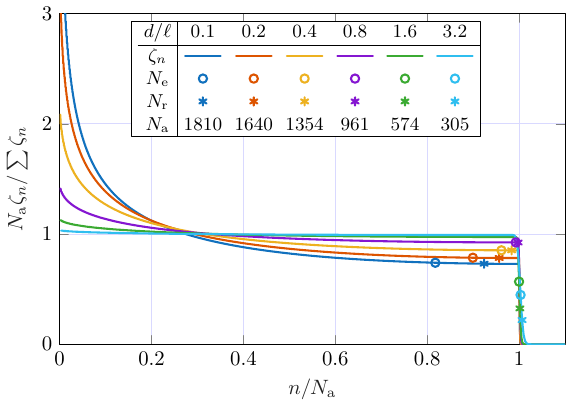}
    \caption{Normalized eigenspectra for two lines with length $\ell$ separated a distance $d$ at wavelength $\lambda/\ell=10^{-3}$, \cf inset in Fig.~\ref{fig:eDoF2lines}.  NDoFs $\Ne$ and $\Nr$ are indicated by the markers and $\Na$ is evaluated from~\eqref{eq:ShadowLength2lines}. }
    \label{fig:linesspectrum}
\end{figure}

The eigenspectra for the channels between the two lines in 3D at $\lambda = 10^{-3}\ell$ are shown in Fig.~\ref{fig:linesspectrum},  with the normalized index axis $n /\Na$. For all $d/\ell$ values, a clear corner (knee) appears around $n=\Na$, beyond ($n>\Na$) which the eigenvalues are negligible, for all separation distances. Markers indicate the effective NDoF~\eqref{eq:eNDoF} and rank~\eqref{eq:rNDoF} estimators.  In the range $n<\Na$, the eigenvalues remain relatively constant only for relatively large separation distances. For these cases, the corner is predicted correctly. For shorter distances, where the NDoF is underestimated (as seen also in Fig.~\ref{fig:eDoF2lines}), a pronounced variation is observed between the leading eigenvalues ($n\approx 1$) and the subsequent ones. Here, it is worth noting that the normalization~\eqref{eq:CouplingStrength} implies a unit area under each curve in Fig.~\ref{fig:linesspectrum}. 
Variations in eigenvalue distribution do manifest, however,  in the denominator of~\eqref{eq:eNDoF}, due to the squaring of the eigenvalues. 
The greater the variation, for the same corner value, the lower the value predicted by~\eqref{eq:eNDoF}.

Similarly, a high-frequency asymptotic expression for the effective NDoF can be derived for parallel lines in $\R^2$. The derivation of such an expression in App.~\ref{S:Lines},  for $\lambda\to 0$,  results in
\begin{equation}
    \Ne^0=
    \frac{\ell}{\lambda}
    \frac{4\beta(\asinh\beta-\sqrt{1+\beta^{-2}}+\beta^{-1})^2}{\frac{2}{3}+\beta\asinh(\beta)
    -\sqrt{1+\beta^2}
    +\frac{1}{3}(1+\beta^2)^{3/2}}.
    \label{eq:eDoF2lines2D}
\end{equation}
The resulting effective NDoF per wavelength $\Ne^0\lambda/\ell$ is shown in Fig.~\ref{fig:eDoF2lines} (dashed yellow line, labeled ``2D"). These results agree with the 3D~\eqref{eq:NeNDoF2lines}, the mutual shadow length~\eqref{eq:ShadowLength2lines}, and paraxial approximation for larger separation distances, \ie $d>2\ell$. For shorter distances, they are the lowest of all estimates and grossly underestimate the corner position. The numerically evaluated $\Ne$ values using~\eqref{eq:eNDoF} approach $\Ne^0$ as the wavelength decreases.

\subsection{Two parallel planar regions}\label{S:TwoPlates}
The asymptotic analysis for the case of parallel lines in Sec.~\ref{S:TwoLines} readily extends to parallel planar regions, as shown in App.~\ref{S:discs}. 
The dominant term in the asymptotic expansion of $\sum\zeta_n^2$ is 
\begin{equation}
    \norm{\M{H}^\T{H}\M{H}}_\T{F}^2
    \approx\frac{\lambda^2}{(4\pi)^4}\int\!\int \frac{R^6}
    {d^2  R^6}\diffS\diffS'
    =\frac{\lambda^2 A_\T{T}A_\T{R}}{(4\pi)^4 d^2}.
    \label{eq:GGplanar}
\end{equation}
The coupling strength normalization term is
\begin{equation}
       \norm{\M{H}}^2_\T{F}
 =\frac{1}{(4\pi)^2}\int\!\int \frac{\diffS\diffS'}
    {(x-x')^2+(y-y')^2+d^2}.
    \label{eq:PlanarCouplingSt}
\end{equation}
 For rectangular regions,  using~\eqref{eq:CouplingStrength2D}, it is reduced to a 2D integral. 
Substituting these expressions into~\eqref{eq:eNDoF} yields the dominant term of the effective NDoF
\begin{equation}
    \Ne^{0}=\frac{\norm{\M{H}}_\T{F}^4 d^2(4\pi)^4}{\lambda^2A_\T{T}A_\T{R}}\leq
    \frac{A_\T{TR}}{\lambda^2},
    \label{eq:eNDoFdiscs}
\end{equation}
where the inequality follows from applying the Cauchy–Schwarz inequality to the coupling strength~\eqref{eq:PlanarCouplingSt}, such that
\begin{equation}
    \norm{\M{H}}_\T{F}^4
    \leq \frac{A_\T{T}A_\T{R}}{(4\pi)^4}
    \int\!\!\int\frac{1}{R^4}\diffS\diffS'
    =\frac{A_\T{T}A_\T{R}}{(4\pi)^4}\frac{A_\T{TR}}{d^2}.
    \label{eq:CS}
\end{equation}
Hence, the effective NDoF $\Ne$ is always bounded by the normalized shadow area $\Na$ for sufficiently short wavelengths. Equality is achieved when the variation of the $R^{-4}$ term is negligible, \ie when the distance between the regions is much larger than their characteristic size, as in the paraxial approximation~\eqref{eq:Aparaxial}.

\begin{figure}
    \centering
    \includegraphics[width=0.95\linewidth]{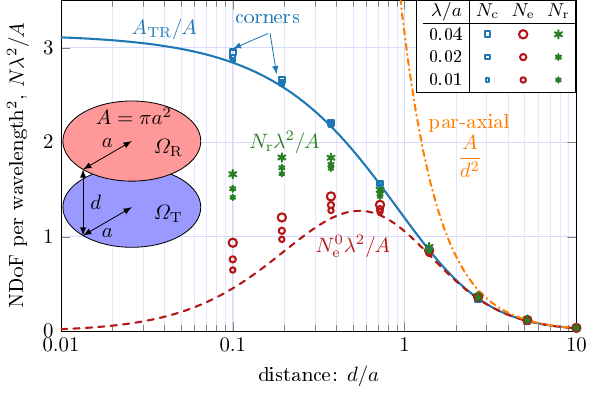}
    \caption{Normalized NDoF for two discs with radius $a$ separated a distance $d$. Numerical results using wavelengths $\lambda/a\in\{0.04,0.02,0.01\}$ are indicated by the markers, with squares indicating corner positions $\Nc$, circles $\Ne$~\eqref{eq:eNDoF}, and asterisks $N_\T{r}$~\eqref{eq:rNDoF}.  The numerical results are compared with analytical expressions for shadow area~\eqref{eq:ShadowA2discs} (solid lines), stationary phase asymptotic $\Ne$ in~\eqref{eq:eNDoFdiscs} (dashed), and the paraxial approximation (dashed dotted).}
    \label{fig:eDoF2discs}
\end{figure}

The asymptotic expression can be calculated for the case of parallel identical discs presented in Fig.~\ref{fig:eDoF2discs}.  In cylindrical coordinates, the coupling strength~\eqref{eq:PlanarCouplingSt} reduces to
\begin{equation}
    \norm{\M{H}}^2_\T{F}=\int_0^r\!\int_0^a\frac{\rho\rho'\diff\rho\diff\rho'}{\sqrt{(\rho+\rho')^2+d^2}\ \sqrt{(\rho-\rho')^2+d^2}},
    \label{eq:DiscsCoupling}
\end{equation}
which can be readily evaluated numerically. 
The dominant term of the effective NDoF, $\Ne^0$, computed using~\eqref{eq:eNDoFdiscs} and~\eqref{eq:DiscsCoupling}, is shown by the dashed curve in Fig.~\ref{fig:eDoF2discs}.  To facilitate the comparison, the values are normalized by $\lambda^2$ and the disc area $A=\pi a^2$. The quantity $\Ne^0$ is small for both short and large separations, with a maximum at intermediate distances. For large separations, $d>a$, it agrees well with the total mutual shadow area~\eqref{eq:ShadowA2discs}. For $d>3a$, it also agrees with the paraxial approximation~\cite{Piestun+Miller2000,Miller2019}. The significant deviation observed for $d<a$ is consistent with the Cauchy–Schwarz inequality~\eqref{eq:CS}, as the $R^{-2}$ term varies considerably in this regime.
The numerical values of $\Ne$, computed for $\lambda \in \{0.04, 0.02, 0.01\}a$, approach $\Ne^0$ as the wavelength decreases, similarly to the case of parallel lines in Fig.~\ref{fig:eDoF2lines}. Also presented are the numerical evaluations of $\Nr$, and corner position, which too coincide for larger separations. The corner positions show excellent agreement with the shadow area-based estimate in the entire range.

\begin{figure}
    \centering
    \includegraphics[width=\linewidth]{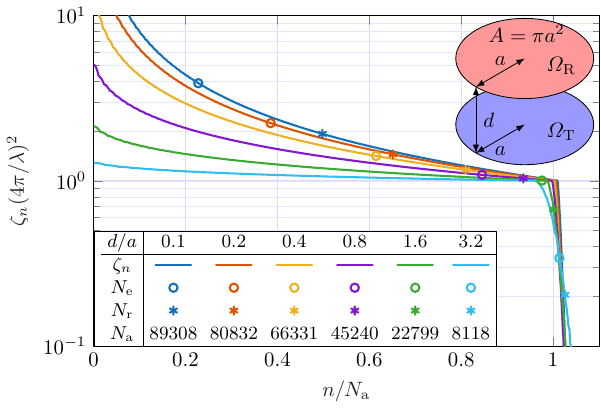}
    \caption{Eigenspectrum for two discs with radius $a$ separated a distance $d$ at wavelength $\lambda=10^{-2}a$ normalized according to~\eqref{eq:averagepropspec}.  NDoFs $\Ne$ and $\Nr$ are indicated by the markers and $\Na$ is evaluated from~\eqref{eq:Na} and~\eqref{eq:ShadowA2discs}, \cf Fig.~\ref{fig:eDoF2discs}. }
    \label{fig:discsspectrum}
\end{figure}

The corresponding eigenspectra for the wavelength $\lambda=0.01a$ and distances $d=\{0.1,0.2,0.4,0.8,1.6,3.2\}a$ are depicted in Fig.~\ref{fig:discsspectrum}. The computed effective NDoF $\Ne$ and effective rank $\Nr$ are indicated by markers. Here too, it is observed that $\Ne\approx\Nr\approx\Na$ for the larger separation distances, where the propagating eigenspectrum is relatively flat. The three predictors differ greatly for shorter distances, where the variation in the propagating eigenspectrum is larger.  

To conclude, for both the lines and flat surface in 3D, as well as in 2D, the deviation of  $\Ne$ from $\Na$ can be predicted analytically for short wavelengths and is indicative of the variation in the propagation spectrum  eigenvalues.

\section{Propagating eigenspectrum distribution}\label{S:ChannelStrength}

Insight into the behavior of the propagating eigenspectrum can be obtained from its average value ~\eqref{eq:averagepropspec}.
By bounding the integrand in the denominator using its minimal and maximal values, we obtain
\begin{multline} 
\frac{1}{\max_{\UV{R}}|\UV{R}\cdot\UV{n}_\T{T}|\ |\UV{R}\cdot\UV{n}_\T{R}|}\\ 
\leq \frac{(4\pi)^2\sum\zeta_n}{\lambda^2\Na} \leq\frac{1}{\min_{\UV{R}}|\UV{R}\cdot\UV{n}_\T{T}|\ |\UV{R}\cdot\UV{n}_\T{R}|}, 
\label{eq:eigvariation}
\end{multline}
where $\UV{R}$ denotes the propagation direction between points in $\regT$ and $\regR$, and $\UV{n}_\T{X}$ is the unit normal of $\reg_\T{X}$. These bounds show that the average level of the propagating eigenspectrum is governed by the incidence angles between the propagation directions and the surface normals at the transmitter and receiver.
Regions for which $|\UV{R}\cdot\UV{n}_\T{X}|$ is small can therefore contribute to larger eigenvalues. Consequently, the average level increases with angular misalignment from the broadside direction, \ie endfire configurations can yield higher eigenvalues than broadside configurations.

In this context, it is worth noting that the case in Fig.~\ref{fig:spheigcomp}, involving a spherical source and a far-field observation sphere, is highly symmetric and provides an extreme example of directional uniformity of all directional beams and observation sectors and, thus, exhibits a rather flat spectrum. Similarly, the much-studied paraxial cases with well-separated parallel regions have low angular diversity and a relatively flat spectrum.

To further understand the variation within the propagating eigenspectrum, we assume that the interaction between the regions can be approximately localized in the electrically large limit, \ie that the overall interaction can be decomposed into contributions from pairs of subregions. This type of asymptotic localization underlies, for example, certain derivations of Weyl’s law. In~\cite{Brick+etal2026}, the propagating spectrum (up to the corner $\Na$) is shown to be associated with directional, beam-producing aperture singular vectors that exhibit localized fields at the observation region.

\begin{figure}
    \centering
    \includegraphics[width=1\linewidth]{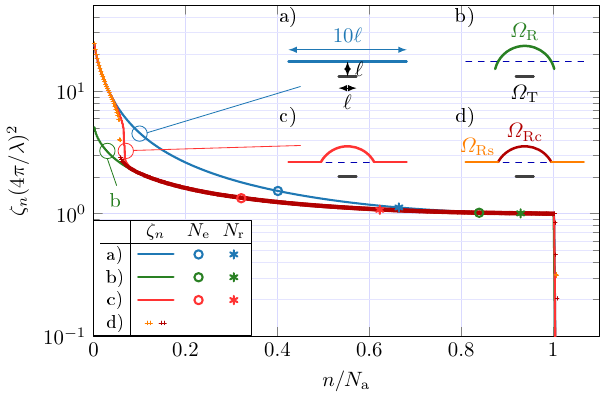}
    \caption{Normalized eigenspectrum for four configurations (a) to (d) with identical $\Na$ evaluated for $\lambda=0.002\ell$. Effective NDoF $\Ne$ and rank NDoF $\Nr$ are indicated by markers. Case (d) is evaluated by splitting $\regR$ into two separate regions $\reg_\T{Rc}$ and $\reg_\T{Rs}$ and merging the eigenvalues in postprocessing.}
    \label{fig:SpectrumLinesArcComp}
\end{figure}

With this in mind, the propagating eigenspectrum is investigated by comparing four configurations with identical mutual shadow area values, as illustrated in Fig.~\ref{fig:SpectrumLinesArcComp}, with the normalized axis $n / \Na$. The two parallel lines in (a) consist of a transmitter of length $\ell$ and a receiver of length $10\ell$, separated by a distance $\ell$. The eigenspectrum for $\lambda = 0.02\ell$ resembles the equal-line case in Fig.~\ref{fig:linesspectrum}, with a few initially very strong eigenvalues followed by approximately equal-strength eigenvalues up to the NDoF $n\approx \Na$. The effective NDoF $\Ne\approx 0.4\Na$ and effective rank $\Nr\approx 0.66\Na$, as indicated by markers in Fig.~\ref{fig:SpectrumLinesArcComp} are positioned in the relatively flat part of the eigenspectrum, relatively far from the corner around $n=\Na$. Comparing the numerically computed eigenvalues with the estimate in~\eqref{eq:eigvariation}, the most oblique propagation direction, occurring between the edges of the regions, yields an upper bound of $31.25$ for the normalized eigenvalues, which is in good agreement with a normalized numerical value for $\zeta_1$ that is roughly $25$ in Fig.~\ref{fig:SpectrumLinesArcComp}.     

Replacing the receiving line by a circular arc, as in Fig.~\ref{fig:SpectrumLinesArcComp}b, while preserving the value for the mutual shadow area, produces a significantly flatter eigenspectrum before the corner at $n \approx \Na$. The arc geometry yields approximately equal distances between the transmitter and the different parts of the receiver and propagation directions approximately normal to $\regR$ for the localized beams produced by the transmitter that illuminate them, \ie $\UV{R}\cdot\UV{n}_\T{R}\approx 1$ in~\eqref{eq:eigvariation}. The angular variation at $\regT$ is the same as in case (a) leading to a bound of approximately $5.6$, which agrees well with the largest eigenvalue $\zeta_1$ of around $5$. This reduces spectral variation across different beam-producing modes and yields a more balanced distribution. 
Consequently, both the effective NDoF $\Ne \approx 0.84\Na$ and the effective rank $\Nr \approx 0.93\Na$ move closer to the asymptotic value $\Na$.

The hybrid geometry in Fig.~\ref{fig:SpectrumLinesArcComp}c exhibits mixed spectral behavior, characterized by a small set of dominant eigenvalues followed by a broad, nearly flat region. Specifically, the spectrum initially resembles configuration (a) for $n < 0.07\Na$, with a few strong modes, and transitions to a flat distribution similar to configurations (b) and (c) for $n > 0.07\Na$. This transition is consistent with a partitioning of the shadow length between the flat and curved segments, \ie $L_\T{TR}/\ell \approx 0.13 + 1.83$, where the flat portion accounts for approximately $7\%$ of the total shadow length. These observations also support the approximate localization perspective: different parts of the interacting regions contribute distinct propagating spectral bands that combine additively for the dominant propagating modes.

The localization observation is further supported by the spectrum for the configuration in Fig.~\ref{fig:SpectrumLinesArcComp}d. There, the receiver region $\regR$ is partitioned into flat shoulders and arc subregions, $\reg_\T{Rs}$ and $\reg_\T{Rc}$. The channels from $\regT$ to each subregion are formed and their eigenvalues are computed separately prior to their sorted merging.
The combined spectrum is shown in yellow and red cross markers for eigenvalues associated with $\reg_\T{Rs}$ and $\reg_\T{Rc}$, respectively. These eigenvalues follow closely those of case (c), with only minor differences, primarily around the transition region near $n \approx 0.07\Na$.
The results in Fig.~\ref{fig:SpectrumLinesArcComp} further indicate that the strongest eigenvalues are primarily associated with the shoulders of the receiving region $\regR$, where propagation angles are largest, rather than with the central region characterized by shorter distances. 
Consequently, the NDoF predicted from the shadow area $\Na$ and those inferred from spectral measures such as $\Ne$ and $\Nr$ may differ in configurations with significant variation in the propagation angle of the interaction.
The comparison also indicates that large-angle interactions primarily determine the largest eigenvalues, whereas the mutual shadow governs the asymptotic position of the corner.

These observations are consistent with the spectral analysis in~\cite{Brick+etal2026}.
The analysis links the largest singular values of the to the singular vectors 
that radiate the beams of the greatest oblique angle seen by the observer. 
For observers significantly wider than the source, as in Fig.~\ref{fig:SpectrumLinesArcComp}a, the beams become localized at regions of the observer and the variation in the singular values increases.

\begin{figure}
    \centering
    \includegraphics[width=1\linewidth]{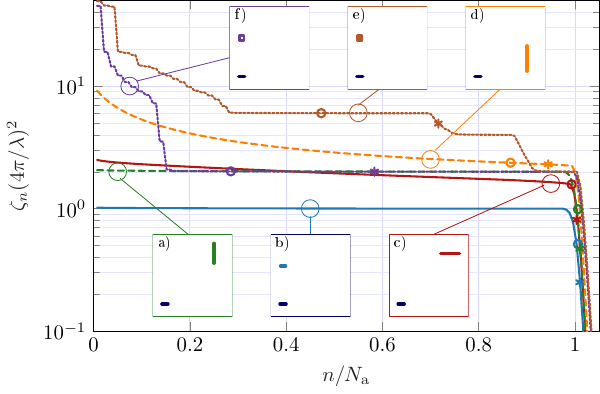}
    \caption{Eigenspectrum for six different setups with similar $\Na\approx 138$ evaluated for $\lambda=0.001\ell$. The average distances $R_0$ are similar for (bdef) and slightly larger for (ac). Effective NDoF $\Ne$ and effective rank $\Nr$ are indicated by circle and star markers, respectively.}
    \label{fig:eigspectrumdiffgeos}
\end{figure}

The last set of examples examines the propagating eigenspectrum for six different configurations (see Fig.~\ref{fig:eigspectrumdiffgeos} insets) with similar mutual shadow lengths. Here, the domains are relatively more distant then in the previous example. In Fig.~\ref{fig:eigspectrumdiffgeos}, star and circle markers indicate $\Ne$ and $\Nr$.  Case (b) with two well-separated parallel lines shows a flat propagating eigenspectrum, as in Fig.~\ref{fig:linesspectrum}. The normalized amplitude is unity in the scaling from~\eqref{eq:averagepropspec} as motivated by the propagation directions normal to the $\regT$ and $\regR$ regions~\eqref{eq:eigvariation}. 
The propagating eigenspectra for cases (a) and (c) are also relatively flat. Their average amplitude, however, is higher because of the oblique directions, following~\eqref{eq:eigvariation}. All these cases have NDoFs $\Ne$ and $\Nr$ close to the corner around $n=\Na$.

Case (d), where the domains are closer together and exhibit greater variation in the range of observation angles and distances, shows a larger variation within the eigenvalues and also a larger average. The resulting NDoFs $\Ne$ and $\Nr$ are now clearly below the corner position.

Cases (e) and (f) have receiving regions consisting of a square in (f) and three inscribed squares in case (e). They have the same visible part as case (b), but extend farther behind the line in (b). Therefore, the shadow area remains unchanged,
while the coupling strength increases due to the larger total support, but with no increase in the angular diversity. However, the NDoFs $\Ne$ and $\Nr$ are well below the corner, with \eg $\Ne\approx\{0.3,0.5\}\Na$.

The examples highlight a trade-off between shadow area, which governs the corner position, and region size and separation, which influence modal strength. In particular, larger visible regions increase the number of available propagating modes.  Larger regions (even if not fully visible to one another) increase the average modal strength. Increased distance reduces the number of propagating modes and the average modal strength.

\section{Conclusions}\label{S:conclusion}
We presented a unified framework for quantifying spatial degrees of freedom and channel strength in radiative electromagnetic systems by combining operator-spectrum and geometry-based analyses. The results indicate that two quantities, mutual shadow (view) area and coupling strength, capture the main characteristics of the propagating eigenvalue spectrum. Specifically, the mutual shadow area determines the spectral-corner location, while coupling strength governs the average level of propagating eigenvalues and thus the overall channel strength.

Using a correlation-operator formulation, we compared spectrum-based NDoF metrics, namely effective NDoF and effective rank, with geometry-based estimates derived from mutual shadow area. The analysis shows that these metrics are consistent for simple configurations under paraxial conditions. For more general geometries, effective rank typically exceeds effective NDoF, while both remain below shadow-based NDoF estimates that track the spectral corner more closely. Asymptotic evaluations for canonical parallel-line and parallel-region configurations in two and three dimensions support these findings and clarify when paraxial interpretations remain accurate.

Differences between spectral and geometric metrics are primarily explained by variability within the propagating eigenvalue spectrum. Configurations with large angular spreads or complex geometries exhibit greater modal variability, which causes effective NDoF and effective rank to underestimate the corner location.

Overall, the proposed framework provides physically interpretable and computationally efficient tools for estimating modal richness and coupling strength in near-field electromagnetic channels. These results are relevant to antenna-array design, electromagnetic inverse problems, integral equations, and high-capacity communication systems, where geometry and propagation distance critically influence performance.

\appendices
\section{Green's functions}\label{S:Green}
This appendix summarizes the Green's functions used in the operator formulation in Section~II and in the asymptotic derivations.
The scalar Green's function in $\R^3$ is
\begin{equation}    
    G_3 = \frac{\exp(-\ju kr)}{4\pi r},
    \label{eq:Green3D}
\end{equation} 
and the Green's dyadic is $\M{G}=(\Id+k^{-2}\nabla\nabla)G_3$.
In $\R^2$, the corresponding scalar Green's function is
\begin{equation}
    G_2 = \frac{\ju}{4}\T{H}^{(2)}_0(kr)
    \approx 
     \frac{1}{4}\sqrt{\frac{2}{\ju\pi kr}}\eu^{-\ju kr}
     =\frac{1}{4\pi}\sqrt{\frac{\lambda}{\ju r}} \eu^{-\ju kr}
     \label{eq:Green2D}
\end{equation}
as $kr\to\infty$, where $\T{H}^{(2)}_0$ denotes a Hankel function.

\section{Derivation for two parallel lines}\label{S:Lines}
This appendix derives the stationary-phase approximation for two parallel lines used in Section~IV.
Consider two parallel lines with length $\ell_1$ and $\ell_2$ separated by a distance $d$. The stationary points are given by the zeros of
\begin{equation}
    \partder{\phi}{x_2}
    =\frac{x_2-x_1'}{\sqrt{(x_2-x_1')^2+d^2}}
    -\frac{x_2-x_2'}{\sqrt{(x_2-x_2')^2+d^2}}
    \label{eq:LineStationary1}
\end{equation}
and similar for $\partder{\phi}{x_2'}$.
This system can be solved by squaring the terms
\begin{equation}
    (x_2-x_1')^2((x_2-x_2')^2+d^2)
    =(x_2-x_2')^2((x_2-x_1')^2+d^2)
\end{equation}
which simplifies to $(x_2-x_1')^2=(x_2-x_2')^2$
and inserting into~\eqref{eq:LineStationary1} $x_2-x_1'= x_2- x_2'$ and $x_1'=x_2'$.
Similarly, from $\partder{\phi}{x_2'}$ we also find $x_1=x_2$.

The Hessian in~\eqref{eq:stationaryphase} at the stationary point follows from $\secondpartder{\phi}{x_2}=0$
and the mixed term
\begin{equation}    
    \frac{\partial^2\phi}{\partial x_2\partial x_2'}
    =\frac{R^2-(x_1-x_1')^2}{R^3}
    =\frac{d^2}{R^3}.
\end{equation}
Thus, the Hessian is
\begin{equation}
    \M{H}_\phi
    =\frac{d^2}{R^3}
    \begin{bmatrix}
        0 & 1\\
        1 & 0
    \end{bmatrix}
    \quad\text{with }
    |\det(\M{H}_\phi)|^{1/2}
    =\frac{d^2}{R^3}
    \label{eq:Hessian}
\end{equation}

\subsection{Lines in 3D}
The phase at the stationary point is $\phi = 0$
giving the stationary phase integral~\eqref{eq:stationaryphase} ($\beta=\ell/d$)
\begin{multline}
    \lambda\int\!\!\int \frac{R^3}
    { d^2 R^4}\diff l_1\diff l_1'
    =\frac{\lambda}{d^2}\int\!\!\int \frac{1}
    {\sqrt{(x_1-x_1')^2+d^2}}\diff l_1\diff l_1'\\
    =\frac{2\lambda}{\ell} (\beta^2\asinh(\beta)-\beta\sqrt{\beta^2+1}+\beta)
\end{multline}
The coupling strength from~\eqref{eq:CouplingStrength} gives
\begin{multline}
       \norm{\M{H}}^2_\T{F}
 =\frac{1}{(4\pi)^2}\int\!\!\int \frac{1}
    {(x_1-x_1')^2+d^2}\diff l_1\diff l_1'\\
    =\frac{2\beta\atan(\beta)-\ln(1+\beta^2)}{(4\pi)^2}
    \label{eq:CouplingStrength2D}
\end{multline}
which yields the dominant asymptotic term in~\eqref{eq:NeNDoF2lines}.

\subsection{Lines in 2D}
The Green's function in 2D~\eqref{eq:Green2D} has the same asymptotic phase as in 3D as $kr\to\infty$ and only differs by the slower amplitude decay.
This gives the same stationary points as for the 3D Green's function in~\eqref{eq:Green3D} and the same Hessian~\eqref{eq:Hessian}. 
The stationary phase integral ($\beta=\ell/d$)
\begin{multline}
    \lambda\int\!\!\int \frac{R^3}
    { d^2 R^2}\diff l_1\diff l_1'
    =\frac{\lambda}{d^2}\int\!\!\int 
    \sqrt{(x_1-x_1')^2+d^2}\diff l_1\diff l_1'\\
    =\frac{2}{3}+\beta\asinh(\beta)
    -\sqrt{1+\beta^2}
    +\frac{1}{3}(1+\beta^2)^{3/2}.
\end{multline}
With normalization from the coupling strength~\eqref{eq:CouplingStrength},
\begin{multline}
       \norm{\M{H}}^2_\T{F}
 =\int\!\!\int \frac{1}
    {((x_1-x_1')^2+d^2)^{1/2}}\diff \ell_1\diff l_1'\\
    =\asinh\beta-\sqrt{1+\beta^{-2}}+\beta^{-1}
    \label{eq:CouplingStrength2Da}
\end{multline}
giving the asymptotic ratio entering the effective-NDoF expression~\eqref{eq:eDoF2lines2D}
and therefore the same linear-in-$\ell/\lambda$ scaling with geometry factor $\beta$ as in the 3D line case.

\section{Derivation for two parallel regions}\label{S:discs}
This appendix outlines the stationary-phase Hessian structure for two parallel regions used in Section~VI.
The stationary points are analogous to the line case in~\eqref{eq:LineStationary1}--\eqref{eq:Hessian}. For the Hessian, the diagonal terms vanish as in the line case, $\secondpartder{\phi}{x_2}=0$. The mixed $x_2$ and  $x_2'$ terms are 
\begin{equation}
    \frac{\partial^2\phi}{\partial x_2\partial x_2'}
    =\frac{1}{R}-\frac{(x_1-x_1')^2}{R^3}
   =\frac{d^2+(y_1-y_1')^2}{R^3}.
\end{equation}
We also need mixed $xy$-derivatives
\begin{equation}    
    \frac{\partial^2\phi}{\partial y_2\partial x_2'}
    =\frac{(x_1-x_1')(y_1-y_1')}{R^3}
    -\frac{(x_1-x_1')(y_1-y_1')}{R^3}
    =0
\end{equation}
and
\begin{equation}    
    \frac{\partial^2\phi}{\partial y_2'\partial x_2'}
    =\frac{(x_1-x_1')(y_1-y_1')}{R^3}.
\end{equation}
Hence the Hessian in~\eqref{eq:stationaryphase} is
\begin{equation}
    \M{H}_\phi=\frac{1}{R^3}
    \begin{bmatrix}
        0 & d^2+y^2 & 0 & xy\\
        d^2+y^2 & 0 & xy & 0 \\
        0 & xy & 0 & d^2+x^2\\
        xy & 0 & d^2+x^2 & 0\\
    \end{bmatrix}
\end{equation}
with the determinant $|\det(\M{H}_\phi)|^{1/2}=d^2R^{-4}$.



\end{document}